\documentclass[12pt]{article}

\oddsidemargin  - 5mm
\evensidemargin - 5mm

\usepackage{amsmath,amssymb}
\usepackage{graphicx}
\usepackage{dcolumn}

\newcommand{\be}{\begin{equation}}
\newcommand{\ee}{\end{equation}}
\newcommand{\bea}{\begin{eqnarray}}
\newcommand{\eea}{\end{eqnarray}}

\newcommand{\sot}{SO(3, \mathbb{R})}
\def\Gr{Gr\"obner}

\begin{document}

\title{
\sc \bf
On application of involutivity analysis of differential equations to constrained
dynamical systems
\footnote{
Talk given at the Seminar
{\em ``Symmetries and Integrable Systems''}, Dubna, Russia, December 20, 2002.}
}
\author{{\em By
Vladimir Gerdt, \,
Arsen Khvedelidze ${}^{a,b}$\,\,
and
Dimitar Mladenov}\\[1cm]
{\small \sf Joint Institute for Nuclear Research, Dubna, Russia}\\
{\small \sf ${}^a$ A.~Razmadze Mathematical Institute, Tbilisi, Georgia}\\
{\small \sf ${}^b$ Department of Mathematics and Statistics,}\\
{\small \sf University of Plymouth, Plymouth, UK}
}
\date{\empty}

\maketitle

\begin{quote}
A brief sketch of computer methods of involutivity analysis of
differential equations is presented in context of its application
to study degenerate Lagrangian systems.
We exemplify the approach by a detailed consideration of a finite-dimensional
model, the so-called light-cone $SU(2)$ Yang-Mills mechanics.
All algorithms are realized in computer algebra system {\em Maple}.
\end{quote}

\renewcommand{\thefootnote}{\fnsymbol{footnote}}
\renewcommand{\thefootnote}{\arabic{footnote}}

\vspace*{1cm}

%%%%%%%%%%%%%%%%%%%%%%%%%%%%%%%%%%% 1 %%%%%%%%%%%%%%%%%%%%%%%%%%%%%%%%%%%%%%%%%%%%

\section{Introduction}

%%%%%%%%%%%%%%%%%%%%%%%%%%%%%%%%%%%%%%%%%%%%%%%%%%%%%%%%%%%%%%%%%%%%%%%%%%%%%%%%%%

Among the properties of systems of analytical partial
differential equations (PDEs) which can be studied without
explicit integration there are two important ones:
the question of their compatibility and the problem of posing of an initial-value
problem (Cauchy problem),
providing the existence and the uniqueness of an analytical solution.
Both of these problems are crucial for the correct formulation of the evolution of
degenerate Lagrangian dynamical systems.

The main obstacle in the study of these problems for PDEs of a
given order $q$ is the existence of ``hidden'' {\em integrability
conditions}. These conditions, $q'\leq q$ order differential
equations, are consequences of the given system of PDEs that can
not be derived using only algebraic manipulations with the PDEs.
The special class of PDEs, called an {\em involutive system} of
PDEs, has all such integrability conditions incorporated in it.
This means that differentiation of the system, called {\em
prolongation}, do not reveal new integrability conditions.
Examples of such involutive systems are the quasilinear systems of
Cauchy-Kowalevskaya type (normal systems).

The extension of a system by its integrability conditions is called a completion.
From the completion point of view the linear homogeneous
systems of PDEs with constant coefficients can be associated with systems of
pure polynomial equations~\cite{Janet}-\cite{G99}.
The polynomial involutive systems~\cite{GB} provide a fruitful algorithmic tool
in commutative algebra~\cite{Seiler2}.

The general algorithmic foundation of the involutive approach is
based on the concept of involutive monomial division invented
in~\cite{GB} and defined for a finite monomial set.
Every particular division provides for each monomial in the set the
self-consistent separation of variables into multiplicative and
non-multiplicative ones.
In the case of linear differential system
all its integrability conditions can be constructed by sequential
performing of multiplicative reductions for non-multiplicative
prolongations of the equations in the system~\cite{G99}.

The polynomial and linear differential involutive systems generate
{\em involutive bases} of polynomial ideals~\cite{GB} and linear
differential ideals~\cite{G99}, respectively.
They are polynomial \cite{BW,CLO} and differential
\Gr\ bases~\cite{Carra,Ollivier} of special form.
Though the involutive bases are generally redundant as the \Gr\ ones,
their use make more accessible the structural
information of the polynomial and differential ideals.
The Janet bases~\cite{GB} may be cited as typical representatives of
involutive bases and have been used in algebraic and Lie symmetry
analysis of differential equations~\cite{G99}.

The completion of differential equations to involution is the most
universal algorithmic method for their algebraic analysis~\cite{G99}
and can be applied for the following purposes:

\begin{itemize}

\item
Check the compatibility of the systems of PDEs.
In case of system inconsistency there is an integrability condition of
the form $1=0$ which is revealed in the course of completion.

\item
Detection of arbitrariness in the general analytic solution of analytic
systems of PDEs~\cite{Pommaret,G99,Seiler3}.

\item
Elimination of a subset of dependent variables, that is, obtaining
differential consequences of the given system, if they exist,
which do not contain the dependent variables specified.

\item
Posing of an initial value problem for a system of analytic PDEs
providing existence and uniqueness of locally holomorphic
solutions~\cite{Pommaret,G99}.

\item
Lie symmetry analysis of differential equations.
Completion to involution of the determining equations for the Lie symmetry generators
is the most general algorithmic method of their integration~\cite{G99,Hereman}.

\item
Preprocessing for numerical integration.
Certain non-commutative involutive bases can be used for automatic generation of
finite-difference schemes PDEs~\cite{BM,MB}.

\item
Computation of the complete set of constraints for degenerated
dynamical systems and their separation into first and second classes
~\cite{Seiler,GG99,G2000}.
\end{itemize}

Below the last application will be exemplified by studying the
finite-dimensional degenerate system, the so-called light-cone
Yang-Mills mechanics. In order to make presentation more
transparent we start with a very short introduction of the main
settings of the involutive method.

%%%%%%%%%%%%%%%%%%%%%%%%%%%%%%%%%%% 2 %%%%%%%%%%%%%%%%%%%%%%%%%%%%%%%%%%%%%%%%%%%%

\section{Involutive polynomial bases}

%%%%%%%%%%%%%%%%%%%%%%%%%%%%%%%%%%%%%%%%%%%%%%%%%%%%%%%%%%%%%%%%%%%%%%%%%%%%%%%%%%

The basic algorithmic ideas go back to M. Janet~\cite{Janet} who
invented the constructive approach to study of PDEs in terms of
the corresponding monomial sets based on the following association
between derivatives and monomials:
\begin{equation}
\frac {\partial^{\mu_1+\cdots+\mu_n} u^\alpha} {\partial
x_1^{\mu_1} \cdots \partial x_n^{\mu_n}} \Longleftrightarrow
[x_1^{\mu_1}\cdots x_n^{\mu_n}]_\alpha\,. \label{d-m}
\end{equation}
Note that the monomials associated to different dependent
variables $u^\alpha$ belong to different monomial sets.

The association (\ref{d-m}) allows to reduce the problem of
involutivity analysis of linear homogeneous systems of PDEs to the
same problem for pure algebraic systems~\cite{Janet}-\cite{G97},\cite{G99}.
Having in mind this fact we shall state now the main definitions and results
concerning the involutivity of algebraic systems.

Let $R=K[x_1,\ldots,x_n]$ be a ring of multivariate polynomials
over a zero characteristic coefficient field $K$.
Then a finite set $F=\{f_1,\ldots,f_m\}\subset R$ of polynomials in $R$ is
{\em a basis of the ideal}
\begin{equation}
<F>=<f_1,\ldots,f_m>=\{\ \sum_{i=1}^m h_if_i\ |\ h_j\in R\ \}\,.
\end{equation}

In the involutive approach to commutative (polynomial)
algebra~\cite{GB}, which is a mapping of the involutivity
analysis of linear PDEs~\cite{Pommaret,G99}, for every polynomial
in the finite set $F$ the set variables ${x_1,\ldots,x_n}$ are
separated into disjoint subsets of {\em multiplicative} and
{\em nonmultiplicative} variables.

To be self-consistent such a separation must satisfy some axioms~\cite{GB} and every
appropriate separation generates an {\em involutive monomial division} in the following sense.
Fix a linear {\em admissible monomial order} $\succ$ such that
\begin{eqnarray}
&& m\neq 1 \Longrightarrow m\succ 1\,, \\
&& m_1\succ m_2\ \Longleftrightarrow m_1m\ \succ m_2m
\end{eqnarray}
holds for any monomials
(power products of the variables with integer exponents) $m$, $m_1$, $m_2$.
Then for every polynomial $f$ in $F$ one
can select its {\em leading monomial} $lm(f)$ (with respect to $\succ$).
All leading monomials in $F$ form a finite monomial set $U$.
If $u\in U$ divides a monomial $w$ such that all the
variables which occur in $w/u$ are multiplicative for $u$, then
$u$ is called {\em involutive divisor} of $w$. We shall denote by
$L$ an involutive division, which specifies a set of
multiplicative (resp. nonmultiplicative) variables for every
monomial $u$ in any given finite monomial set $U$ and write $u|_Lw$ if $u$ is
($L-$)involutive divisor of $w$.
In the latter case we shall also write $w=u\times v$ where, by the above definition,
the monomial $v=w/u$ contains only multiplicative variables.

In the papers~\cite{GB,G98,Yufu} several involutive divisions were
introduced and studied in detail.
Here, as an example we present one called after M. Janet,
who was one of the founders of the involutive approach to PDEs and
who devised the related separation of variables~\cite{Janet}.

Given a finite set $U$ of monomials in $\{x_1,\ldots,x_n\}$ and a
monomial $u=x_1^{d_1}\cdots x_n^{d_n}\in U$, a variable $x_i$
$(i>1)$ is Janet multiplicative for $u$ if its degree $d_i$ in $u$
is maximal among all the monomials in $U$ having the same degrees
in the variables $x_1,\ldots,x_{i-1}$.
As for $x_1$, it is {\em Janet multiplicative} for $u$ if $d_1$ takes the maximal value among
the degrees in $x_1$ of monomials in $U$.
If a variable is not Janet multiplicative for $u$ in $U$ it is considered as
{\em Janet nonmultiplicative}.

Consider, for example, a monomial set
\begin{equation}
U=\{x_1x_2,x_2x_3,x_3^2\}\,.
\end{equation}
This gives the following Janet multiplicative and nonmultiplicative
variables for monomials in $U$:

\begin{center}
\vskip 0.2cm
\begin{tabular}{|c|c|c|} \hline\hline
Monomial &\multicolumn{2}{c|}{Variables}
 \\ \cline{2-3}
 &   Multiplicative        & Nonmultiplicative  \\ \hline
$x_1x_2$ &   $x_1,x_2,x_3$ & $-$   \\
$x_2x_3$ &   $x_2,x_3$     & $x_1$    \\
$x_3^2$  &   $x_3$         & $x_1,x_2$  \\ \hline
\hline
\end{tabular}
\vskip 0.2cm
\end{center}

Given a finite polynomial set $F$, a noetherian~\cite{GB}
involutive division $L$, for instance, Janet division, and an
admissible monomial order $\succ$, one can algorithmically
construct~\cite{GB} a {\em minimal $L-$invol\-utive basis} or
$L-$basis $G\subset R$ of the ideal $<F>=<G>$ such that for any
polynomial $f$ in the ideal there is a polynomial $g$ in $G$
satisfying $lm(g)|_L lm(f)$, and every polynomial $g$ in $G$ does
not contain monomials having involutive divisors among the
leading monomials of other polynomials in $G$.

If $F=\{f_1,\ldots,f_m\}\subset R$ is a polynomial set, $L$ is an
involutive division and $\succ$ is an admissible monomial order,
then any polynomial $p$ in $R$ can be rewritten (reduced) modulo
the ideal $<F>$ as
\begin{equation}
p=h-\sum_{i=1}^m \sum_j a_{ij}f_i\times u_{ij}\,,
\end{equation}
where $a_{ij}$ are elements (coefficients) of the base field
$K$, $u_{ij}$ are $L-$multiplicative monomials for $lm(f_i)$ such
that $lm(f)\,u_{ij}\preceq lm(p)$ for all $i,j$, and there are no
monomials occurring in $h$ which have $L-$involutive divisors
among $\{lm(f_1),\ldots,lm(f_m)\}$. In this case $h$ is said to
be in the {\em $L-$normal form} modulo $F$ and written as
$h=NF_L(p,F)$.

If $G$ is $L-$basis, then $NF_L(p,G)$ is uniquely defined
%___________________________________________________________________%
\footnote{
For other properties of the involutive bases,
proofs and illustrating examples see~\cite{GB}.}
%____________________________________________________________________%
for any polynomial $p$.
In this case $NF_L(p,G)=0$ if and only if $p$ belongs to the ideal $<G>$
generated by $G$. Moreover, if the ideal is {\em radical} for
which any its element (polynomial) vanishes at the common roots
of all the polynomials in $G$ if and only if this polynomial
belongs to the ideal, then it follows that the condition
$NF_L(p,G)=0$ is necessary and sufficient for vanishing $p$ on
those common roots.

It is important to emphasize that any involutive basis is a \Gr\
basis, generally redundant, and can be used in the same manner as
the reduced \Gr\ basis~\cite{BW,CLO}.

The above described and some other properties of the involutive
bases as well as the \Gr\ bases allow one to work fully
algorithmically~\cite{GG99,G2000} with constraints in the case of
degenerated dynamical systems of polynomial type. In particular,
one can work algorithmically on the constraint surface. In the
next section considering the light-cone Yang-Mills mechanics we
shall show how the above mentioned ideas of the involutive
analysis can be realized in the computer algebra system {\em Maple}.

%%%%%%%%%%%%%%%%%%%%%%%%%%%%%%%%%%% 3 %%%%%%%%%%%%%%%%%%%%%%%%%%%%%%%%%%%%%%%%%%%%

\section{Application to Yang-Mills light-cone mechanics}

%%%%%%%%%%%%%%%%%%%%%%%%%%%%%%%%%%%%%%%%%%%%%%%%%%%%%%%%%%%%%%%%%%%%%%%%%%%%%%%%%%

Before demonstration of computer calculations we formulate the
Yang-Mills light-cone mechanics.
At first we use the standard Dirac-Hamilton formalism for systems with constraints.
Then we explain some computational aspects of deriving the same results
implementing the Dirac's method in terms of involutive polynomial
bases~\cite{G2000} based on the Maple package.

%##################################### 3.1 #########################################%

\subsection{Dirac's constrained dynamics}

%###################################################################################%

The Yang-Mills mechanics was formulated twenty years ago as
instant form Yang-Mills field theory with spatially constant gauge
fields and has been intensively studied during the last decades
from different standpoints (see e.g \cite{MatSav}-\cite{KM} and references therein).
The light-cone version of Yang-Mills mechanics is formulated analogously,
it follows from the classical action functional for Yang-Mills field theory with an Ansatz that
the gauge potential is a function of the light-cone time only.
So, we start with the action for the pure Yang-Mills gauge field
in four-dimensional Minkowski space $M_4$, endowed with a metric $\eta$
%________________________________________________________________________________%
\footnote{
In this paper we follow the notations of Ref. \cite{VAD1,VAD2}}
\begin{equation}
\label{eq:gaction}
I : = \frac{1}{g^2}\, \int_{M_4} \mbox{tr} \,  F\wedge * F \,,
\end{equation}
where $g$ is a coupling constant and the $su(2)$ algebra valued curvature two-form
\begin{equation}
F:= d A +  A \wedge A
\end{equation}
is constructed from the connection one-form $A$.
The connection and curvature, as Lie algebra valued quantities, are expressed in terms
of the antihermitian $su(2)$ algebra basis $\tau^a = \sigma^a/2 i$
with the Pauli matrices $\sigma^a \,, a = 1,2,3$,
\begin{equation}
A = A^a \, \tau^a \,, \qquad F = F^a \, \tau^a\,\,.
\end{equation}
The metric $\eta$ enters the action through the dual field strength
tensor defined in accordance to the Hodge star operation
$
* F_{\mu\nu}  = \frac{1}{2}\,\sqrt\eta\, \epsilon_{\mu\nu\alpha\beta}\, F^{\alpha\beta}\,.
$
The light-cone coordinates $ x^\mu = \left( x^+, x^-, x^\bot\right)$ are chosen as
\begin{equation}
x^\pm := \frac{1}{\sqrt 2}\, \left( x^0 \pm x^3 \right) \,, \qquad
x^\bot :=  \left( x^k \,, k = 1, 2 \right)
\end{equation}
and the non-zero components of the metric $\eta$ in the light-cone basis are
\begin{equation}
\eta_{+-} = \eta_{-+} = - \eta_{11} = - \eta_{22} = 1\,.
\end{equation}
The connection one-form in the light-cone formulation is given as
\begin{equation} \label{eq:conlc}
A = A_+ \, dx^+ + A_- \, dx^- + A_k \, dx^k \,.
\end{equation}

By definition the Lagrangian of the light-cone Yang-Mills mechanics follows from
the corresponding Lagrangian of Yang-Mills theory if one supposes that
connection one-form $A$ depends on the light-cone ``time variable''  $x^+$ alone
\begin{equation}
A = A(x^+) \,.
\end{equation}
Using the definitions (\ref{eq:gaction}) and (\ref{eq:conlc})
we find the Lagrangian of the Yang-Mills light-cone mechanics
\begin{equation} \label{eq:lagr}
L : = \frac{1}{2 g^2} \,
\left(
F^a_{+ -} \,  F^a_{+ -} + 2 \, F^a_{+ k} \, F^a_{- k} - F^a_{12} \, F^a_{12}
\right)\,,
\end{equation}
where the light-cone components of the field-strength tensor are given by
\begin{eqnarray}
&& F^a_{+ -} = \frac{\partial A^a_-}{\partial x^+} + \epsilon^{abc}\, A^b_+ \,  A^c_- \,,\\
&& F^a_{+ k} = \frac{\partial A^a_k}{\partial x^+} + \epsilon^{abc}\, A^b_+ \,  A^c_k \,, \\
&& F^a_{- k} = \epsilon^{abc} \,  A^b_- \, A^c_k \,, \\
&& F^a_{i j} = \epsilon^{abc}\, A^b_i \, A^c_j\,, \quad i,j,k = 1,2 \,.
\end{eqnarray}

Choosing the light-cone coordinate $x^+$ as an evolution parameter $\tau$
\begin{equation}
\tau := x^+
\end{equation}
we define the light-cone version of the Hamiltonian description of a system with Lagrangian
(\ref{eq:lagr}).
The corresponding Hessian is degenerate, namely
$\mbox{corank}||\frac{\partial^2 L}{\partial \dot{A}\partial \dot{A}}||= 6$,
and thus the Legendre transformation
%_____________________________________________________________________________%
\footnote{
To simplify the formulas we shall use overdot to denote
derivative of functions with respect to light-cone time  $\tau$.
Further, we shall treat in equal footing the up and down
isotopic indexes denoted with $a, b, c, d$.}
%______________________________________________________________________________%
\begin{eqnarray}
&&\pi^+_a  =  \frac{\partial L}{\partial \dot{A^a_+}} =0\,,\\
&&\pi^-_a  =  \frac{\partial L}{\partial \dot{A^a_-}} =
\frac{1}{g^2} \, \left( \dot{A^a_- } + \epsilon^{abc} \, A^b_+ \, A^c_- \right) \,, \\
&& \pi_a^k = \frac{\partial L}{\partial \dot{A^a_k}} =
\frac{1}{g^2} \, \epsilon^{abc} \, A^b_- \, A^c_k \,,
\end{eqnarray}
impose the six primary constraints on the canonical coordinates.
Thus the generalized Hamiltonian system is characterized besides the
canonical Poincar\'{e}-Cartan one-form
\begin{eqnarray}\label{Poincare-Cartan}
\Theta_C: = \pi^+_a\ dA^a_+  + \pi^-_a\ dA^a_- + \pi_a^i\ dA^a_i -H_C\ d\tau
\end{eqnarray}
by set of primary constraints
\begin{eqnarray}
&&
\varphi^{(1)}_a := \pi^+_a = 0 \,, \label{eq:prcon-1} \\
&&
\chi^a_k := g^2 \, \pi^a_k  + \epsilon^{abc} \, A^b_- A^c_k =
0\,,\label{eq:prcon-21}
\end{eqnarray}
satisfying the following Poisson brackets relations
\begin{eqnarray}
&& \{ \varphi^{(1)}_a \,, \varphi^{(1)}_b\} = 0 \,,\\
&& \{ \varphi^{(1)}_a \,, \chi^b_k \} = 0 \,,\\
&& \{ \chi^a_i \,, \chi^b_j\} = -2\, g^2 \epsilon^{abc}\, A^c_- \, \eta_{i j} \,.\label{eq:prcon-2}
\end{eqnarray}
In (\ref{Poincare-Cartan}) the canonical light-front Hamiltonian is
\begin{equation} \label{eq:khrlh}
H_C = \frac{g^2}{2}\,  \pi^-_a  \pi^-_a - \,
\epsilon^{abc} \,  A^b_+ \left(A^c_- \, \pi^-_a  + A^c_k \,\pi^k_a \right) + V(A_k)\,,
\end{equation}
where the potential term $V$ is
\begin{equation}
V(A_k) = \frac{1}{2 g^2} \,
\left[
\left(A^b_1 A^b_1\right)\, \left(A^c_2 A^c_2 \right) -
\left(A^b_1 A^b_2\right)\, \left(A^c_1 A^c_2 \right)
\right]\,.
\end{equation}

Following the Dirac formulation, the dynamics of the
light-cone Yang-Mills mechanics is governed by the total Hamiltonian
\begin{equation} \label{eq:toth}
H_T = H_C - 2 \, \mbox{tr} \left(U (\tau)\, \varphi^{(1)} \right) -
2 \, \mbox{tr}\left(V_k(\tau)\, \chi_k \right)\,,
\end{equation}
where $U(\tau)$ and $V_k(\tau)$ are arbitrary $SU(2)$ valued
functions of the light-cone time $\tau$.
Using this Hamiltonian we
find that there are three secondary constraints $\varphi^{(2)}_a$
\begin{equation} \label{eq:secgauss}
\varphi^{(2)}_a := \epsilon_{abc}
\left(A^b_-  \pi^-_c  +  A^b_k \pi^k_c \right)=0\,,
\end{equation}
obeying  the $so(3, \mathbb{R})$ algebra
\begin{equation}
\{ \varphi^{(2)}_a \,, \varphi^{(2)}_b \} = \epsilon_{abc}\, \varphi^{(2)}_c \,.
\end{equation}
Checking analogously the time evolution of  the primary constraints
$\chi^a_k$ we have
\begin{equation} \label{eq:secchi}
0 = {\dot\chi}^a_k =
\{\chi^a_k\,, H_C \} - 2\, g^2\, \epsilon^{abc} \, V^b_k\, A^c_-   \,.
\end{equation}
The analysis of this equation depends on the properties of
the matrix $\mathcal{C}_{ab} = \epsilon^{abc}\, A^c_-$.
This matrix is degenerate with a rank vary
from $0$ to $2$ depending on the point of the configuration space.
If its rank is $2$ then among the six primary constraints
$\chi^a_k$ there are two first class constraints and maximum four
Lagrange multipliers $V$ can be determined from (\ref{eq:secchi}).
When the rank of the matrix $\mathcal{C}_{ab}$ is minimal, the locus points are
$A^a_-=0$ and all six constraints $\chi^a_k$ are Abelian ones.
For such an exceptional configuration the constrained system
reduces to dynamically trivial one and hereinafter
we shall consider the subspace of configuration space where
${\rm rank}||\mathcal{C}||=2$.
For such configurations we are able to introduce the unit vector
\begin{equation}
N^a = \frac{A^a_-}{ \sqrt{(A_-^1)^2 + (A_-^2)^2 + (A_-^3)^2} }\,,
\end{equation}
that is a null vector of the matrix
$ \| \, \epsilon^{abc} \, A^c_- \,\|,$
and to rewrite  the six primary constraints $\chi^a_k$ as
\begin{eqnarray}
&& \chi^a_{k \bot} :=
\chi^a_k - \left( N^b \chi^b_k  \right) \, N^a \,, \\
&& \psi_k : = N^a \chi^a_k  \,.\label{eq:abpsi}
\end{eqnarray}
Constraints $\chi^a_{k \bot}$ are functionally dependent due to the conditions
\begin{equation} \label{eq:dependence}
N^a \, \chi^a_{k \bot} = 0\,
\end{equation}
and choosing among them any four independent constraints,
we are able to determine four Lagrange multipliers $V^k_{\ b \bot}$.
The two constraints $\psi_k$ satisfy the Abelian algebra
\begin{equation}
\{ \psi_i \,, \psi_j \} = 0 \,.
\end{equation}

The Poisson brackets of the constraints $\psi_k$ and $\varphi^{(2)}_a$ with the total Hamiltonian
vanish after projection on the constraint surface (CS)
\begin{eqnarray} \label{eq:check}
&&
\{ \psi_k \,, H_T \}{\,\vert_{CS}} = 0 \,, \\
&&
\{ \varphi^{(2)}_a \,, H_T \}{\,\vert_{CS}} = 0\,
\end{eqnarray}
and thus there are no ternary constraints.

Summarizing, we arrive at the set of constraints
$ \varphi^{(1)}_a, \psi_k, \varphi^{(2)}_a, \chi^b_{k \bot}$.
The Poisson bracket algebra of the three first ones is
\begin{eqnarray}
&& \{ \varphi^{(1)}_a \,, \varphi^{(1)}_a\} = 0 \,, \label{eq:group_1} \\
&& \{ \psi_i \,, \psi_j \} = 0 \,, \label{eq:group_2} \\
&& \{ \varphi^{(2)}_a \,, \varphi^{(2)}_b\} = \epsilon_{abc}\, \varphi^{(2)}_c \,,\label{eq:group_3} \\
&& \{ \varphi^{(1)}_a \,, \psi_k\} =  \{\varphi^{(1)}_a \,,\varphi^{(2)}_b \} =
\{ \psi_k \,, \varphi^{(2)}_a \} = 0 \,. \label{eq:group_4}
\end{eqnarray}
The constraints $\chi^b_{k \bot}$ satisfy the relations
\begin{equation} \label{eq:bracket-1}
\{ \chi^a_{i \bot} \,, \chi^b_{j \bot} \} = - 2 \, g^2 \, \epsilon^{abc} \, A^c_- \, \eta_{i j}\,,
\end{equation}
and the Poisson brackets between these two sets of constraints are
\begin{eqnarray}
&&
\{\varphi^{(2)}_a \,, \chi^b_{k \bot} \} =
\epsilon^{abc} \, \chi^c_{k \bot} \,, \label{eq:bracket-2}\\
&& \{ \varphi^{(1)}_a \,, \chi^b_{k \bot}\} =
\{ \psi_i \,, \chi^b_{j\bot}\} = 0 \,. \label{eq:bracket-3}
\end{eqnarray}

From these relations we conclude that we have 8 first-class constraints
$\varphi^{(1)}_a, \psi_k, \varphi^{(2)}_a $
and 4 second-class constraints $\chi^a_{k \bot}$.
This means that now constraints reduce the $24$ constrained phase space
degrees of freedom to  $24 - 4 - 2 (5 + 3) = 4$
unconstrained degrees of freedom,
in contrast to the instant form of the Yang-Mills mechanics
where the number of the unconstrained canonical variables is $12$.

%%%%%%%%%%%%%%%%%%%%%%%%%%%%%%%%%%% 4 %%%%%%%%%%%%%%%%%%%%%%%%%%%%%%%%%%%%%%%%%%%%

\subsection{Computational aspects of Dirac-\Gr\ algorithm}

%%%%%%%%%%%%%%%%%%%%%%%%%%%%%%%%%%%%%%%%%%%%%%%%%%%%%%%%%%%%%%%%%%%%%%%%%%%%%%%%%%

Now we shall discuss what kind of computer algebra manipulations
are necessary to perform in order to obtain the above stated results.
We shall follow the general algorithm \cite{GG99} adapted
to the computer algebra manipulations in theories with polynomial Lagrangians.
This algorithm, called Dirac-\Gr\ algorithm, combines
the constructive ideas of Dirac formalism for constrained systems
with the \Gr\ bases techniques.

Denote by $q_\alpha$ and $\dot{q}_\alpha$,
$1\leq \alpha \leq 12$, respectively, the generalized Lagrangian
coordinates in~(\ref{eq:lagr}) listed as
\begin{equation}
A^1_{+},A^2_{+},A^3_{+},A^1_1,A^2_1,A^3_1,A^1_2,A^2_2,A^3_2,A^1_-,A^2_-,A^3_-
\end{equation}
and their velocities (time derivatives).
Then the momenta are
$p_\alpha=\frac{\partial L}{\partial \dot{q}_\alpha}\,, 1\leq \alpha \leq 12$.
To compute the primary constraints it suffices
to eliminate the velocities $\dot{q}_\alpha$ from these system treated
as polynomial in $\dot{q}_\alpha,q_\alpha,p_\alpha$.
The elimination is performed by computing the \Gr\ basis~\cite{BW,CLO}
for the generating polynomial set
\begin{equation}
\{\ p_\alpha-\frac{\partial L}{\partial \dot{q}_\alpha}\ \mid \ 1\leq \alpha\leq 12\ \}
\end{equation}
for an ordering (in Maple{\tt lexdeg}) eliminating velocities
$\dot{q}_\alpha$.
In the obtained set all algebraically dependent constraints~\cite{CLO} are ruled out.

The canonical Hamiltonian~(\ref{eq:khrlh}) is determined as reduction of
\begin{equation}
p_\alpha\dot{q}_\alpha - L
\end{equation}
modulo the \Gr\ basis computed.
Then computation of the Poisson brackets between the constraints is straightforward.

Next step is the construction of the secondary
constraints~(\ref{eq:secgauss}). To find them we  reduce the
Poisson brackets of the primary constraints with the total
Hamiltonian modulo the set of primary constraints. Again the \Gr\
basis technique provides the right algorithmic tool for doing such
computations and to obtain a complete set of twelve algebraically
independent constraints. In order to compare with the constraints
given in the previous section, we represent them as
\be\label{eq:complete}
 \mathcal{F}=\{\ \varphi^{(1)}_a, \psi_k,
\varphi^{(2)}_a, \chi^1_{k}, \chi^2_{k} \ \}\,, \qquad a,b=1,2,3;\
k=1,2\,,
\ee
Next, to separate the complete set of constraints
into first and second classes we  compute the $12\times 12$
Poisson bracket matrix on the constraint surface
\begin{equation}
M:=\|\, \{f_\alpha\,,f_\beta\}_{\,\vert{CS}} \,\| \,,
\qquad \alpha, \beta = 1, \dots 12\,,
\end{equation}
where $f_\alpha,f_\beta\in \mathcal{F}$.
Since the $\mbox{rank}\| M \| = 4$ the complete constraint set
$\mathcal{F}$ can be separated to four second-class constraints
and eight first-class ones.
To select the first-class constraints it suffices to compute the basis
$\mathcal{A}=\{\mathbf{a}_1,\ldots,\mathbf{a}_8\}$
of the null space of the matrix $\| M \|$ and then construct the first-class
constraints as
\begin{equation}
First~~class~~constraints\ = \
\sum_{\alpha=1}^{12}(\mathbf{a}_i)_\alpha f_\alpha\,,
\qquad 1\leq i\leq 8\,.
\end{equation}
To extract the second-class constraints from $\mathcal{F}$ one constructs $8\times 12$ matrix
$\|\, (\mathbf{a}_i)_\alpha\, \|$ from the components of the
vectors in $\mathcal{A}$ and find the basis
$\mathcal{B}=\{\mathbf{b}_1,\ldots,\mathbf{b}_4\} $
of the null space of the constructed matrix.
Then every vector $\mathbf{b} \in \mathcal{B}$ yields a second-class constraint
\begin{equation}
Second~~class~~constraints =
\sum_{\alpha=1}^{12}(\mathbf{b}_i)_\alpha f_\alpha,
\qquad 1\leq i\leq 4\,.
\end{equation}

As a result, one can organize the eight first-class constraints as
$\varphi^{(1)}_a, \psi_k, \varphi^{(2)}_a $, whereas the four
algebraically independent constraints $\chi^1_{k},\chi^2_{k}$ are
second-class.

Relations (\ref{eq:group_1})-(\ref{eq:group_4}) revealing the
structure of the gauge group generated by the first class
constraints can also be computed fully algorithmically. To do this
we extended the Maple package~\cite{GG99} with a general procedure
to represent the Poisson brackets of any two first-class
constraints $f_\alpha$ and $f_\beta$ as a linear combination of
elements in the set of first-class constraints:
\be
\label{eq:gauge_group} \{f_\alpha\,,f_\beta\} = c^\gamma_{\ \alpha
\beta}\, f_\gamma \,.
\ee
With that end in view and in order to cope the most general case we implemented
the extended \Gr\ basis algorithm~\cite{BW}.
Given a set of polynomials $Q=\{q_1,\ldots,q_m\}$ generating the polynomial ideal $<Q>$, this
algorithm outputs the explicit representation
\be \label{eq:gb}
g_\alpha=h_{\alpha \beta}\,q_\beta
\ee
of elements in the \Gr\ basis $G=\{g_1\ldots,g_n\}$ of this ideal in terms of
the polynomials in $Q$.
Having computed the \Gr\ basis $G$ for the ideal generated by
the first-class constraints and the corresponding polynomial
coefficients $h_{\alpha \beta}$ for the elements in $G$ as given
in~(\ref{eq:gb}), the local group coefficients $c^\gamma_{\ \alpha
\beta}$ (which may depend on the generalized coordinates and
momenta) in~(\ref{eq:gauge_group}) are easily computed by
reduction~\cite{BW,CLO} of the Poisson brackets modulo \Gr\ basis
expressed in terms of the first-class constraints.

However, the use of this universal approach may be very expensive
from the computational point of view.
For this reason our Maple package tries first to apply the multivariate polynomial division
algorithm~\cite{CLO} modulo the set of first-class constraints.
Due to the special structure of the primary first-class
constraints that usually include those linear in momenta as
in~(\ref{eq:prcon-1}), this algorithm often produces the right
representation~(\ref{eq:gb});
but unlike the extended \Gr\ basis algorithm does it very fast.
Correctness of the output is easily verified by computing of the reminder.
If the last vanishes, then the output of the division algorithm is correct.
Otherwise the extended \Gr\ basis algorithm is applied.

In our case the division algorithm just produces the correct
formulas (\ref{eq:group_1})-(\ref{eq:group_4}) for the Poisson
brackets of the first-class constraints
$\varphi^{(1)}_a, \psi_k, \varphi^{(2)}_a$.
Similarly, one obtains the formulas (\ref{eq:bracket-1})-(\ref{eq:bracket-3}).

%%%%%%%%%%%%%%%%%%%%%%%%%%%%%%%%%%% 5 %%%%%%%%%%%%%%%%%%%%%%%%%%%%%%%%%%%%%%%%%%%%

\section{The unconstrained system as conformal mechanics}

%%%%%%%%%%%%%%%%%%%%%%%%%%%%%%%%%%%%%%%%%%%%%%%%%%%%%%%%%%%%%%%%%%%%%%%%%%%%%%%%%%

Now following the paper \cite{VAD1} we demonstrate how using
Hamiltonian reduction of the degrees of freedom one can derive the
unconstrained version of the light-cone $SU(2)$ Yang-Mills
mechanics which coincides with the well-known model, the so-called
conformal mechanics \cite{deAlfaroFubiniFurlan}.

To show this equivalence we rewrite the model in terms of special
coordinates as follows.
At first we organize the configuration
variables $A^a_i$ and $A^a_-$ in $3\times 3$ matrix $A_{ab}$ whose
entries of the first two columns are $A^a_i$ and the third column is
composed by the elements $A^a_-$
\begin{equation}
A_{ab} : = \| A^a_1\,, A^a_2\,, A^a_-\|.
\end{equation}
In order to find an explicit parameterization of the orbits and
the slice structure with respect to the gauge symmetry action, it
is convenient to use a polar decomposition  for the matrix
$A_{ab}$
\begin{equation}\label{eq:polar}
A = O S\,,
\end{equation}
where $S$ is a positive definite $3 \times 3$ symmetric matrix,
$O(\phi_1,\phi_2, \phi_3) = e^{\phi_1 J_3}e^{\phi_2 J_1}e^{\phi_3
J_3}$ is an orthogonal matrix parameterized by the three Euler
angles $(\phi_1,\phi_2,\phi_3)$ and $\sot$ generators in adjoint
representation $(J_a)_{ij} = \epsilon_{iaj}$. Here we assume that
the matrix $A$ has a positive determinant and  hence treat the
polar decomposition (\ref{eq:polar}) as a uniquely invertible
transformation from the configuration variables $A_{ab}$ to a new
set of Lagrangian variables: six coordinates $S_{ij}$ and three
coordinates $\phi_i$.
%_____________________________________________________________________________________%
\footnote{
Note that the polar decomposition is valid for an arbitrary
matrix but the orthogonal matrix in
(\ref{eq:polar}) is uniquely determined only from the invertible matrix
$O = A S^{-1 }\,\,, S=\sqrt{A A^T}~$
and thus only in this case the polar decomposition (\ref{eq:polar})
is a well-defined coordinate transformation.
Concerning the degenerate matrices the more close and subtle
analysis should be done.
Here we note only that the set of $n \times n$ matrices
with rank $k$ is a manifold with dimension $k(2n-k)$, but as distinct from
the non-degenerate case now the manifold atlas contains with a necessity several charts.}
%______________________________________________________________________________________%
The polar decomposition (\ref{eq:polar}) induces the point
canonical transformation from the coordinates
$A_{ab}$ and $ \Pi_{ab}:= \| \pi^{a1}\,, \pi^{a2}\,, \pi^{a-} \|$
to new canonical pairs
$(S_{ab}, P_{ab})$ and $(\phi_a, P_a)$ with the following  non-vanishing
Poisson brackets
\begin{eqnarray}
&&
\{ S_{ab} \,, P_{cd} \} =
\frac{1}{2}
\left( \delta_{ac}\, \delta_{bd} + \delta_{ad} \, \delta_{bc} \right)\,,\\
&&
\{ \phi_{a} \,, P_{b} \} =  \delta_{ab}\,.
\end{eqnarray}
The expression of the old $\Pi_{ab}$ as a function of the new coordinates is
\begin{equation}
\Pi = O \left( P - k_a J_a \right)\,,
\end{equation}
where
\begin{equation}
k_a = \gamma^{-1}_{ab} \left(\eta^L_b - \varepsilon_{bmn}\left(S P \right)_{mn}  \right)\,,
\end{equation}
$
\gamma_{ik} = S_{ik} - \delta_{ik}\, \mbox{tr} S
$
and $\eta^L_a$ are three left-invariant vector fields on $\sot$
group (the explicit form of $\eta^L_a$ in terms of the angular variables
$\phi_{a} \,, P_{a}$ can be found in \cite{VAD1}).
In terms of the new variables the
constraints  $\varphi^{(2)}_a$ and $\chi^a_i$ can be rewritten in
the equivalent  form
\begin{eqnarray}\label{eq:gc}
&&
\eta^L_a = 0\,, \\
&&
\widetilde{\chi}_{ai} = P_{ai} + \epsilon_{aij}\,\gamma^{-1}_{jk}\ \epsilon_{kmn}(SP)_{mn}
+ \epsilon_{amn}\, S_{m3}\, S_{ni}=  0\,
\end{eqnarray}
with vanishing Poisson brackets
\begin{equation}\label{eq:nalg}
\{\eta^L_a, \widetilde{\chi}_{bi}\} = 0\,.
\end{equation}

Passing to the new polar decomposition variables we achieve the complete separation of variables
$(S_{ab},P_{cd})$, which are invariant under the gauge transformations generated by
Gauss law constraints $\varphi^{(2)}_a$, from the gauge variant one $(\phi_a, P_a)$.
Owing to the separation to eliminate all gauge degrees of freedom related to
this symmetry it is enough to project to the constraint shell
described by the condition of nullity of Killing fields $\eta_a^L$.
The corresponding pure gauge degrees of freedom will automatically
disappear from the projected Hamiltonian.

In order to proceed further in elimination of remaining constraints  we introduce
the main-axes decomposition for the symmetric $3\times 3$ matrix $S$
\begin{equation}\label{eq:mainax}
S =
R^T(\chi_1, \chi_2, \chi_3)
\left(
\begin{array}{ccc}
q_1   &   0    &    0 \\
0     &   q_2  &    0 \\
0     &   0    &    q_3
\end{array}
\right)
R(\chi_1, \chi_2, \chi_3)\,,
\end{equation}
with an orthogonal matrix $R$, parameterized by three Euler angles $\chi_1,\chi_2,\chi_3$.
Because the Jacobian of this transformation is
$
J\left(\frac{S[q, \chi]}{q, \chi} \right) \sim \prod_{a \neq
b}^{3} \mid q_a - q_b \mid
$
the equation (\ref{eq:mainax}) can be
used as definition of new configuration variables: three
``diagonal'' variables $(q_1, q_2, q_3)$, eigenvalues of the
matrix $S$, and three angular variables $(\chi_1, \chi_2,\chi_3)$,
if and only if all eigenvalues of the matrix $S$ are different.
To have the uniqueness of the inverse transformation we assume here that
$
q_1 < q_2 < q_3\,.
$
%______________________________________________________________________________________________%
\footnote{
The variables $q_a$ in the main-axes transformation
(\ref{eq:mainax}) parameterize the orbits of the adjoint action of
$\sot$ group in the space of $3 \times 3$ symmetric matrices.
Whereas the consideration of the configuration $q_1 < q_2 < q_3$
describing the so-called principle orbit class given below is
correct, the treatment of all orbits with coinciding eigenvalues
of the matrix $S$, the singular orbits, requires more skillful
treatment that is beyond the scope of the present paper.}
%_______________________________________________________________________________________________%

The momenta $p_a$ and $p_{\chi_a}$, canonically conjugated
to the diagonal and angular variables $q_a$ and $\chi_a$,
can be found using the condition of the canonical invariance of the symplectic one-form
\begin{equation}
\sum^3_{a, b = 1}\, P_{ab} \, d S_{ab} =
\sum^3_{a = 1} \, p_a\, d q_a  + \sum^3_{a = 1}\, p_{\chi_a}\, d {\chi}_a \,.
\end{equation}

The original momenta $P_{ab}$, expressed in terms of the new canonical variables,
read
\begin{eqnarray} \label{eq:newmom}
P = R^T \sum_{a = 1}^3 \left( p_a \, \overline{\alpha}_a -
\frac{1}{2}\, \frac{\xi^R_a}{q_b - q_c} \,\alpha_a \right)
R\,,\,\,\, (\mbox{cyclic permutations} \, a\not = b \not = c)\,.
\end{eqnarray}
Here $\overline{\alpha}_a$ and $\alpha_a$ denote the diagonal and
off-diagonal basis elements for the space of symmetric matrices
which obey the relations
$
\mbox{tr}\
(\overline{\alpha}_a \overline{\alpha}_b) = \delta_{ab},
\quad \mbox{tr}\ ({\alpha}_a\, {\alpha}_b) = 2 \delta_{ab},\\
\quad \mbox{tr}\ (\overline{\alpha}_a {\alpha}_b)~=~ 0.
$

The $\xi^R_a $ are three $\sot$ right-invariant vector fields
given in terms of the angles $\chi_a$ and their conjugated momenta $p_{\chi_a}$ via
\begin{eqnarray}
&&
\label{eq:rf1}
\xi^R_1 =
 - \sin\chi_1 \cot\chi_2 \ p_{\chi_1} +
\cos\chi_1 \  p_{\chi_2} +
\frac{\sin\chi_1}{\sin\chi_2}\ p_{\chi_3} \,,\\
&&
\label{eq:rf2}
\xi^R_2 =
\,\,\cos\chi_1 \cot\chi_2 \ p_{\chi_1} + \sin\chi_1 \  p_{\chi_2} -
\frac{\cos\chi_1}{\sin\chi_2}\ p_{\chi_3} \,, \\
&&
\label{eq:rf3}
\xi^R_3 =  p_{\chi_1}\,.
\end{eqnarray}
The constraints $\widetilde{\chi}$ rewritten in the main-axes variables take the form
\begin{eqnarray} \label{eq:mcons}
\widetilde{\chi} = \sum_{a = 1}^3\, R^T
\left[
\pi_a \, \overline{\alpha}_a - \frac{1}{2}\, \rho^-_a \alpha_a +
\frac{1}{2}\, \rho^+_a\, J_a \right] R\,,
\end{eqnarray}
where
\begin{equation}
\rho^\pm _a = \frac{\xi^R_a}{q_b\ \pm q_c} \pm \frac{1}{g^2}\, q_a n_a(q_b\ \pm \ q_c)\,,
\end{equation}
and $n_a = R_{a3}$.

As it was shown above the constraints $\chi^a_i$ represent a mixed system of first
and second class constraints, $\psi_i$ and ${\chi^a_i}_\bot$ correspondingly.
To perform the reduction to the constraint shell it is useful at first to introduce
a gauge fixing condition and  eliminate the two first class constraints $\psi_i$.
The expression (\ref{eq:abpsi}) for the Abelian constraints $\psi_i$ dictates the
appropriate gauge fixing condition
\begin{equation}\label{eq:gcon}
\overline{\psi}_i := N^a A^a_i = 0\,,
\end{equation}
which is canonical one in the sense that
$
\{\overline{\psi}_i,\ \psi_j\} \ =\  \delta_{ij}\,.
$
The constraints  $\psi_i=0 $ together with the canonical
gauge-fixing condition (\ref{eq:gcon}) rewritten in terms of the
main-axis variables fixes the canonical angular variables
\begin{equation}\label{eq:magf}
\chi_1 = 0\,,\ \ \ p_{\chi_1} = 0\,, \qquad \chi_2 = {\pi\over
2}\,,\ \ p_{\chi_2} = 0 \,.
\end{equation}

Examine the remaining four second class constraints $\chi^1_i$ and
$\chi^2_i$ in terms of the main-axes variables we find that the
corresponding constraint shell can be described by the following
conditions on the ``diagonal'' canonical pairs
\begin{eqnarray}\label{eq:sma1}
p_1 =0 \,, \quad p_3=0\,,\quad  (q_1\pm q_3)^2 \ =\ \pm \ g^2\
 \frac{\xi_2^R}{q_2} \,.
\end{eqnarray}
Now using all above expressions for the constraints
one can easily project the total Hamiltonian (\ref{eq:toth})
on the constraint shell and
convince that the dynamics of the two unconstrained canonical pairs $(q_2, p_2)$ and
$(\chi_3, p_{\chi_3})$ is governed by the following reduced Hamiltonian
\begin{equation}\label{eq:hymcm}
  H_{LC}^\ast\ = \ \frac{g^2}{2}\left( p^2_{2}\  + \frac{\alpha}{q_2 ^2}\right)\,,
\end{equation}
where $\alpha = \frac{p_{\chi_3}^2 }{4} \,.$ \footnote{Here we use
that the expression for $\xi_2^R$ reduces to $-p_{\chi_3}$ on the
constraint shell.} Because  the momentum $p_{\chi_3}$ is conserved
one can identify the reduced Hamiltonian of $SU(2)$ light-cone
mechanics with the  Hamiltonian of conformal mechanics whose
``coupling constant'' is determined by the value of $\alpha$,
while the gauge field coupling constant $g$ controls the scale for
the evolution parameter.

\section{Acknowledgments}
This work was supported in part by the RFBR
grant 01-01-00708. Contribution of V.G. was also partially
supported by grant 2339.2003.2 from the Russian Ministry of Industry, Science and
Technologies. A.K. acknowledges INTAS for providing financial
support, grant 00-00561.


\begin{thebibliography}{99}
%
% 1
\bibitem{Janet}
M. Janet,
{\bf Le\c cons sur les Syst\`emes
d'Equations aux D\'eriv\'ees Partielles},
Cahiers Scientifiques, IV, Gauthier-Villars, Paris, 1929.
%
% 2
\bibitem{Pommaret}
J.F. Pommaret,
{\bf Partial Differential Equations and Group Theory. New Perspectives for Applications},
Kluwer, Dordrecht, 1994.
%
% 3
\bibitem{G97}
V.P. Gerdt,
{\it Gr\"obner bases and involutive methods for algebraic and differential equations},
Math. Comp. Model. {\bf 25} (1997) 75-90.
%
% 4
\bibitem{GB}
V.P. Gerdt and Yu. A. Blinkov,
{\it Involutive bases of polynomial ideals},
Mathematics and Computers in Simulation {\bf 45} (1998) 519-542;
{\it Minimal involutive bases}, ibid., 543-560.
%
% 5
\bibitem{G99} V.P. Gerdt,
{\it Completion of linear differential systems to involution},
In: "Computer Algebra in Scientific Computing  / CASC'99",
Springer-Verlag, Berlin, 1999, pp. 115-137.
%
% 6
\bibitem{Seiler2}
W.M. Seiler,
{\it A combinatorial approach to involution and delta-regularity I,
II: Structure analysis of polynomial modules with Pommaret bases},
Preprint Universit\"{a}t Mannheim 2002.
%
% 7
\bibitem{BW}
T. Becker, V. Weispfenning, and H Kredel,
{\bf\it Gr\"obner Bases. A Computational Approach to Commutative Algebra},
Graduate Texts in  Mathematics {\bf 141}, Springer-Verlag, New York, 1993.
%
% 8
\bibitem{CLO}
D. Cox, J. Little, and D. O'Shea,
{\bf Ideals, Varieties and  Algorithms},
2nd Edition, Springer-Verlag, New York, 1996.
%
% 9
\bibitem{Carra}
G. Carra'Ferro,
{\it Gr\"obner bases and differential algebra},
Lec. Not. in Comp. Sci. {\bf 356} (1987) 129-140.
%
% 10
\bibitem{Ollivier}
F. Ollivier,
{\it Standard bases of differential ideals},
Lec. Not. in Comp. Sci. {\bf 508} (1990) 304-321.
%
% 11
\bibitem{G98}
V.P. Gerdt,
{\it Involutive division technique:
some generalizations and optimizations},
Journal of Mathematical Sciences {\bf 108(6)} (2002) 1034-1051.
%
% 12
\bibitem{Yufu}
Chen Yu-Fu and Gao Xiao-Shan,
{\it Involutive directions and new involutive divisions},
Comp. and Math. with Appl. {\bf 41} (2001) 945-956.
%
% 13
\bibitem{Seiler3}
W.M. Seiler,
{\it Arbitrariness of the general solution and symmetries},
Acta Applicandae Mathematicae {\bf 41} (1995) 311-322.
%
% 14
\bibitem{Hereman} H
W. Hereman,
{\it Symbolic software for the computation of Lie symmetry analysis},
In:  CRC Handbook of Lie Group Analysis of Differential Equations,
Volume 3: {\bf New Trends in Theoretical Developments and Computational methods},
N.M. Ibragimov et al. (eds.),
CRC Press, Boca Raton, 1995, pp. 367-413.
%
% 15
\bibitem{BM}
Yu.A. Blinkov and V.V. Mozzhilkin,
{\it Finite volume method for higher-order equations},
In: {\bf Aerodynamics. Nonlinear Problems},
Saratov State University, 1997, pp.140-148 (in Russian).
%
% 16
\bibitem{MB}
V.V. Mozzhilkin and Yu. A. Blinkov,
{\it Methods for constructing finite difference schemes in gas dynamics},
In: Proceedings of Saratov State University,
Vol. 1 (2001) 145-156 (in Russian).
%
% 17
\bibitem{Seiler}
W.M. Seiler and R.W. Tucker,
{\it Involution and constrained dynamics},
J. Phys. A. {\bf 28} (1995) 4431-4451.
%
% 18
\bibitem{GG99}
V.P. Gerdt and S.A. Gogilidze,
{\it Constrained Hamiltonian systems and Gr\"{o}bner bases},
In: Computer Algebra in Scientific Computing \/ CASC'99,
V.G. Ganzha, E.W. Mayr and E.V. Vorozhtsov (Eds.),
Springer-Verlag, Berlin, 1999, pp. 138-146.
%
% 19
\bibitem{G2000}
V.P. Gerdt,
{\it Computer algebra and constrained dynamics},
In: Problems of Modern Physics, A.N. Sisakian and D.I. Trubetskov (Eds.),
JINR D2-99-263, 2000, pp. 164-171.
%
% 20
\bibitem{MatSav}
S.G. Matinyan, G.K. Savvidy and N.G. Ter-Arutyunyan-Savvidy,
{\it Classical Yang-Mills mechanics. Nonlinear colour oscillations},
Sov. Phys. JETP {\bf 53} (1981) 421-429.
%
% 21
\bibitem{AsatSav}
H. M. Asatryan and G.K. Savvidy,
{\it Configuration manifold of Yang-Mills classical mechanics},
Phys. Lett. {\bf A 99} (1983) 290-292.
%
% 22
\bibitem{Medvedev}
B.V. Medvedev,
{\it Dynamic stochasticity and quantization},
Theor. Math. Phys. {\bf 60} (1985) 782-797.
%
% 23
\bibitem{MSolov}
M.A. Soloviev,
{\it On the geometry of classical mechanics with nonabelian gauge symmetry},
Teor. Math. Phys. {\bf 73} (1987) 3-15.
%
% 24
\bibitem{DahRaab}
B. Dahmen and B. Raabe,
{\it Unconstrained $SU(2)$ and $SU(3)$ Yang-Mills classical mechanics},
Nucl. Phys. {\bf B 384} (1992) 352-380.
%
%  25
\bibitem{DYMM}
S. A. Gogilidze, A.M. Khvedelidze, D.M. Mladenov, and  H.-P. Pavel,
{\it Hamiltonian reduction of $SU(2)$ Dirac-Yang-Mills mechanics},
Phys. Rev. {\bf D 57} (1998) 7488-7500.
%
% 26
\bibitem{KM}
A.M. Khvedelidze and D.M. Mladenov,
{\it Euler-Calogero-Moser system from $SU(2)$ Yang-Mills theory},
Phys. Rev. {\bf D 62} (2000) 125016(1-9).
%
% 27
\bibitem{VAD1}
V.P. Gerdt, A.M. Khvedelidze, and D.M. Mladenov,
{\it Light-cone $SU(2)$ Yang-Mills theory and conformal mechanics},
[arXiv: hep-th/0210022].
%
% 28
\bibitem{VAD2}
V.P. Gerdt, A.M. Khvedelidze, and D.M. Mladenov,
{\it Analysis of constraints in light-cone version of $SU(2)$ Yang-Mills mechanics},
In: Proceedings of International Workshop Computer Algebra and its Application to Physics,
Dubna, 2001, \,[arXiv: hep-th/0209107].
%
% 29
\bibitem{deAlfaroFubiniFurlan}
V. De Alfaro, S. Fubini, and G. Furlan,
{\it Conformal invariance in quantum mechanics},
Nuovo Cimento {\bf 34} (1976) 569-612.
\end{thebibliography}
\end{document}